\documentclass[prb, amsmath, amssymb, reprint]{revtex4-2}

\usepackage{graphicx}
\usepackage{dcolumn}
\usepackage{bm}
\usepackage[utf8]{inputenc}
\usepackage[T1]{fontenc}
\usepackage{mathptmx}

\begin{document}

\preprint{APS/123-QED}

\title[Excitation and Detection of THz Coherent Spin Waves in Antiferromagnetic $\alpha$-Fe$_2$O$_3$]{Excitation and Detection of THz Coherent Spin Waves in Antiferromagnetic $\alpha$-Fe$_2$O$_3$}

\author{K. Grishunin}
\email{k.grishunin@science.ru.nl}
\altaffiliation[Also at ]{MIREA—Russian Technological University, Moscow 119454, Russia}

\author{E.A. Mashkovich}
\affiliation{Radboud University, Ultrafast Spectroscopy of Correlated Materials, Nijmegen 6525AJ, The Netherlands}
\author{A.M. Balbashov}
\affiliation{Moscow Power Engineering Institute, Moscow 111250, Russia}
\author{A.K. Zvezdin}
\affiliation{Prokhorov General Physics Institute of the Russian Academy of Sciences, Moscow 119991, Russia}
\author{A.V. Kimel}
\affiliation{Radboud University, Ultrafast Spectroscopy of Correlated Materials, Nijmegen 6525AJ, The Netherlands}

\date{\today}

\begin{abstract}
The efficiency of ultrafast excitation of spins in antiferromagnetic $\mathrm{\alpha-Fe_{2}O_{3}}$ using nearly single-cycle THz pulse is studied as a function of the polarization of the THz pulse and the sample temperature. Above the Morin point the most efficient excitation is achieved when the magnetic field of the THz pulse is perpendicular to the antiferromagnetically coupled spins. Using the experimental results and equations of motion for spins, we show that the mechanism of the spin excitation above and below the Morin point relies on magnetic-dipole interaction of the THz magnetic field with spins and the efficiency of the coupling is proportional to the time derivative of the magnetic field.
\end{abstract}

\maketitle
\section{Introduction}
In quantum mechanics, the Coulomb interaction between two electrons depends on the mutual orientations of their spins. This short-range spin-dependent part of the Coulomb interaction, also known as the exchange interaction, is able to induce a long-range ferromagnetic and antiferromagnetic order of spins, respectively \cite{Neel1948,Dzyaloshinsky1958}. Although materials with ferromagnetic order (ferromagnets) are used in conventional magnetic data storage, spintronics and magnonics technologies, antiferromagnets represent the largest, the least explored and probably the most intriguing family of magnetically ordered materials in nature.

It is believed that antiferromagnets can dramatically improve the performance of these technologies in terms of densities and speed \cite{Piramanayagam2011,Gomonay2014,Olejnik2018}. The absence of net magnetization and stray fields eliminates crosstalk between neighbouring bits or devices \cite{Piramanayagam2011}, and the absence of primary macroscopic magnetization makes the spin manipulation in antiferromagnets inherently faster than in ferromagnets \cite{Gomonay2014,Thielemann-Kuhn2017}. Fundamental studies of spin dynamics, the developments of means and approaches for manipulation and detection of spins in antiferromagnets are presently among the hottest topics in magnetism \cite{Jungwirth2016,Lebrun2020,Baltz2018,Lebrun2020}. 

Recently, it was argued that nearly single cycle terahertz (THz) pulses are the most energy efficient means of ultrafast spin control in antiferromagnets. Several reports demonstrated an efficient excitation of spins in antiferromagnetics NiO \cite{Kampfrath2011,Baierl2016b}, $\mathrm{TmFeO_3}$ \cite{Baierl2016c,Grishunin2018,Schlauderer2019} and $\mathrm{FeBO_3}$ \cite{Mashkovich2019}. However, it is surprising that despite several attempts no THz excitation and detection of coherent spin waves has been reported for hematite - the main component of rust, the most widespread mineral in nature and, obviously, the most widespread antiferromagnet.

Here we explore the efficiency of THz excitation of spins in antiferromagnetic $\mathrm{\alpha-Fe_{2}O_{3}}$. We demonstrate experimentally and theoretically that the mechanism of spin excitation relies on magnetic-dipole interaction of the THz magnetic field and spins and the efficiency of the coupling is proportional to the time derivative of the THz magnetic field.

The paper is organized as follows. Chapter II reports about details of the studied sample, its characterization and experimental procedure. Chapter III describes the main experimental results. Chapter IV proposes a thermodynamic theory describing possible modes of spin resonance in this compound, temperature dependencies of the modes and torques which can excite them. Based on comparison of the theory with the experimental results, the chapter draws conclusions which are summarized in the last section.

\section{Experimental procedure}
Hematite $\mathrm{\alpha-Fe_2O_3}$ is a prominent representative of a broad class of canted antiferromagnetic iron-oxides. The $ \mathrm{Fe^{3+}}$ ions form two magnetic sublattices, the spins of which are antiferromagnetically (AF) coupled due to the symmetric exchange interaction.  Similarly to $\mathrm{FeBO_3}$, crystal structure of hematite $\mathrm{\alpha-Fe_2O_3}$ has $\mathrm{\overline{3}m}$ point group \cite{Dzyaloshinsky1958,Turov1965}.  

Below the Morin temperature ($T<T_M=260$K), the magnetic moments are oriented along the [001] crystallographic axis and the material has no net magnetization. At $T_{\mathrm{M}}$ hematite undergoes a spin-flop phase transition. Above the Morin point, the direction of spins is parallel to the basal (001) plane with a slight in-plane canting. This canted or weak ferromagnetic (WF) phase persists up to the Néel temperature ($T_N=950$ K) above which the material becomes paramagnetic \cite{Morin1950}. For the first time, the existence of the WF state was observed experimentally in hematite more than 60 years ago \cite{Morin1950} and explained by Dzyaloshinskii \cite{Dzyaloshinsky1958} and Moriya \cite{Moriya1960} as the consequence of an antisymmetric exchange interaction between atomic spins. It should be also noted that near the Morin transition temperature the WF and AF phases may coexist \cite{Dang1998}.

\begin{figure*}
\includegraphics[width=\textwidth]{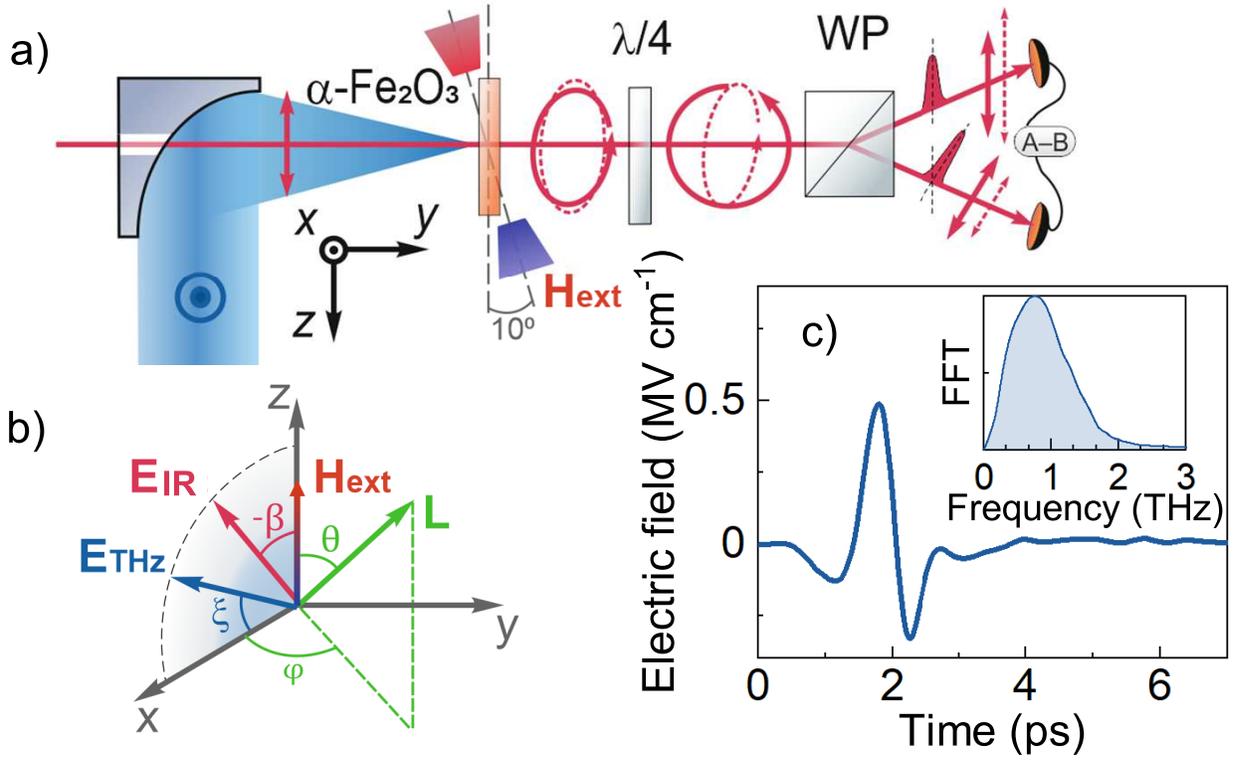}
\caption{\label{fig:geometry} Experimental geometry. (a) Experimental schematic for ultrafast terahertz (THz) pump–infrared (IR) probe spectroscopy. The corresponding polarization transformations of the IR pulse in the absent and the present of THz are shown by solid and dash lines, respectively; (b) The corresponding schematic representation of the mutual orientations of antiferromagnetic vector $\mathbf{l}$, the electric field of the THz ($E_{THz}$) and IR ($E_{IR}$) pulses and the external magnetic field $\mathbf{H_{ext}}$. (c) The incident THz pulse in the time domain and it's Fourier transform (insert).}
\end{figure*}

In our studies we used a 500-$\mathrm{\mu m}$ thick $\mathrm{\alpha}$-hematite ($\mathrm{\alpha-Fe_{2}O_{3}}$) with [001] crystallographic axis along the normal to the sample. The crystal was grown by floating zone melting method under oxygen pressure of 60 atm at the temperature 1400 C with a growth rate of 8 mm/h. The details of the crystal growing can be found elsewhere \cite{Balbashov2019}. The high optical crystal quality as well as the crystallographic orientation were confirmed by IR transparent and X-ray topography.

We excited $\mathrm{\alpha-Fe_{2}O_{3}}$ with intense nearly single-cycle terahertz pulses generated by tilted-pulse-front optical rectification \cite{Hebling2002} of near infrared laser pulses from the amplified Ti:Sapphire laser ($\lambda=800$ nm; pulse duration $\Delta\tau=100$ fs, repetition rate is 1 kHz).
The generated THz pulse beam was collimated and focused onto the sample by using off-axis parabolic mirrors (see Fig. \ref{fig:geometry}a). The focal lengths of the mirrors were chosen to provide the smallest spot diameter about of 500 $\mathrm{\mu m}$ at FWHM. The polarization of the THz pulse was varied by using a combination of two wire-grid polarizers where the second polarizer set the output THz polarization direction, while the first ensured constant intensity ($I_{max}/4$) onto the sample. The peak THz electric field was up to 0.5 MV/cm. The pulse spectrum range from 0.1 to 2 THz covers the characteristic frequency of the antiferromagnetic resonance (Fig.\ref{fig:geometry}c). The induced ultrafast spin dynamics is revealed by tracking the ellipticity of a co-propagating 100 fs IR laser pulse generated by optical parametric amplifier with the central wavelength $\lambda=1350$ nm in the transmission geometry. To ensure that after every pump pulse and relaxation the very same initial state of the medium is restored, the external magnetic field ${H_{\mathrm{ext}}\approx 0.64 \mathrm{ kOe}}$ was applied at an angle of 10 degrees to the sample plane.

Conventionally, the magnetic structure of weak ferromagnets is described by  antiferromagnetic  $\mathbf{l} = (\mathbf{M_1}-\mathbf{M_2})/2M_0$ and ferromagnetic $\mathbf{m} = (\mathbf{M_1}+\mathbf{M_2})/2M_0$ macrospin parameters, where $M_1$ and $M_2$ are the saturation magnetizations of the two antiferromagetically coupled sublattices and ${M_0 = 870\mathrm{Gs}}$ is the saturation magnetization for each of the sublattices.

\begin{figure}
\includegraphics{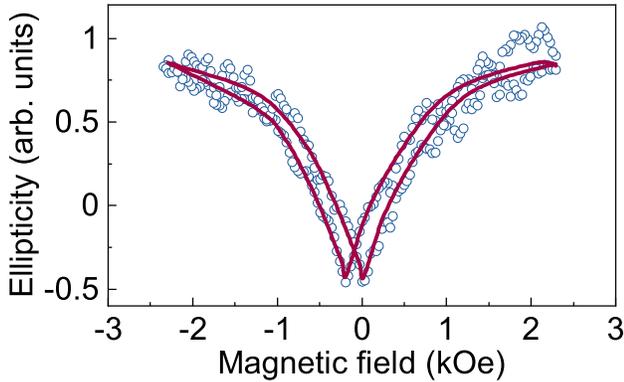}
\caption{\label{fig:hysteresis_ellipticity} Ellipticity acquired
by linearly polarized light as a function of the external magnetic field at $T=293$ K. The solid line serves as a guide of the eye.}
\end{figure}

Figure \ref{fig:hysteresis_ellipticity} shows how the ellipticity acquired by linearly polarized light upon propagation through the sample depends on the applied magnetic field at room temperature. The dependence is even with respect to the applied external magnetic field which corresponds to the modulation of the symmetric part of the dielectric tensor. The saturation does not occur and the magnetization will continue to grow until the external magnetic field becomes comparable with the field of the exchange interaction.

\section{Results}
\begin{figure*}
\includegraphics[width=\textwidth]{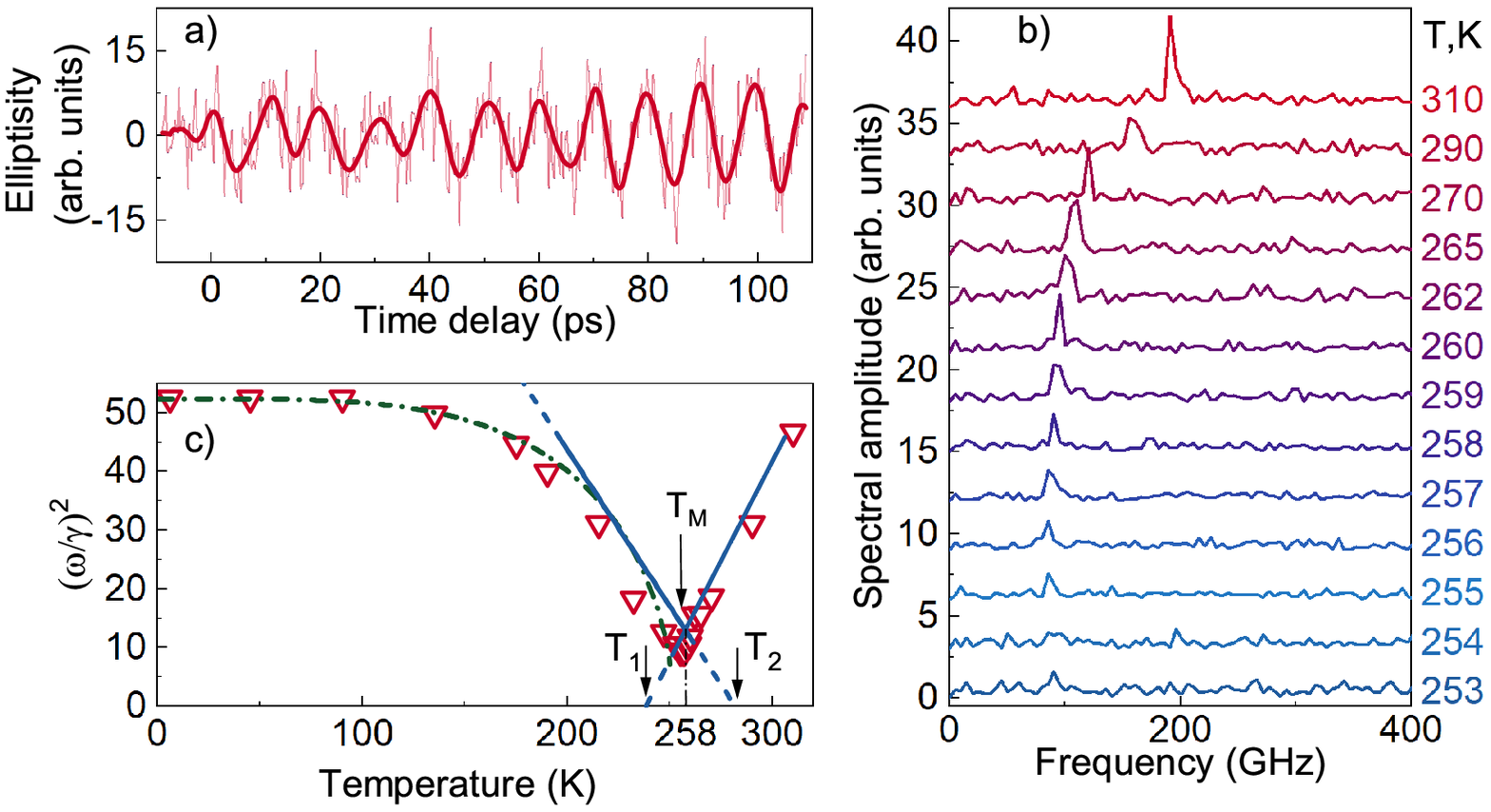}
\caption{\label{fig:temp_dep} Terahertz-induced magnetization dynamic in $\mathrm{\alpha-Fe_{2}O_{3}}$. (a) The dependence of elliptisity of the IR pulse on the time delay between THz pump and IR probe pulse. The solid line is a filtered signal, serving as a guide for the eye; (b) Amplitude spectra of the magnon traces for various sample temperatures; (c) The resonance frequencies of the q-AFM mode as a function of the sample temperature. Red triangles show the experimental data; blue lines show the fit with Eqs. \ref{eq:AFM_frequency} and \ref{eq:AFM_frequency_LowT}; a green dash dot line shows a fit with a spin-wave model \cite{Nagai1975,Chou2012}}
\end{figure*}

A time trace of the THz-induced dynamics contains oscillations with the frequency of $\omega\approx200$ GHz at room temperature (Fig. \ref{fig:temp_dep}a). The frequency of the observed oscillations is in the ballpark of the one for the antiferromagnetic (q-AFM) mode in this compound. We performed the time-resolved measurements in a broad range of temperatures and processed the obtained time traces using the Fourier transform (see Fig. \ref{fig:temp_dep}b). It is seen that both the amplitude and the frequency of the oscillations change dramatically with temperature. Close to the temperature of the Morin transition, the frequency of oscillations softens down to $\omega\approx 82$ GHz. With a further decrease of the temperature to 6K, the frequency slowly increases and returns to $\approx200$ GHz (Fig.\ref{fig:temp_dep}c). This behavior is in qualitative agreement with the one expected for the q-AFM mode in the hematite \cite{Velikov1969,Chou2012,Mikhaylovskiy2015b}.

\begin{figure}
\includegraphics{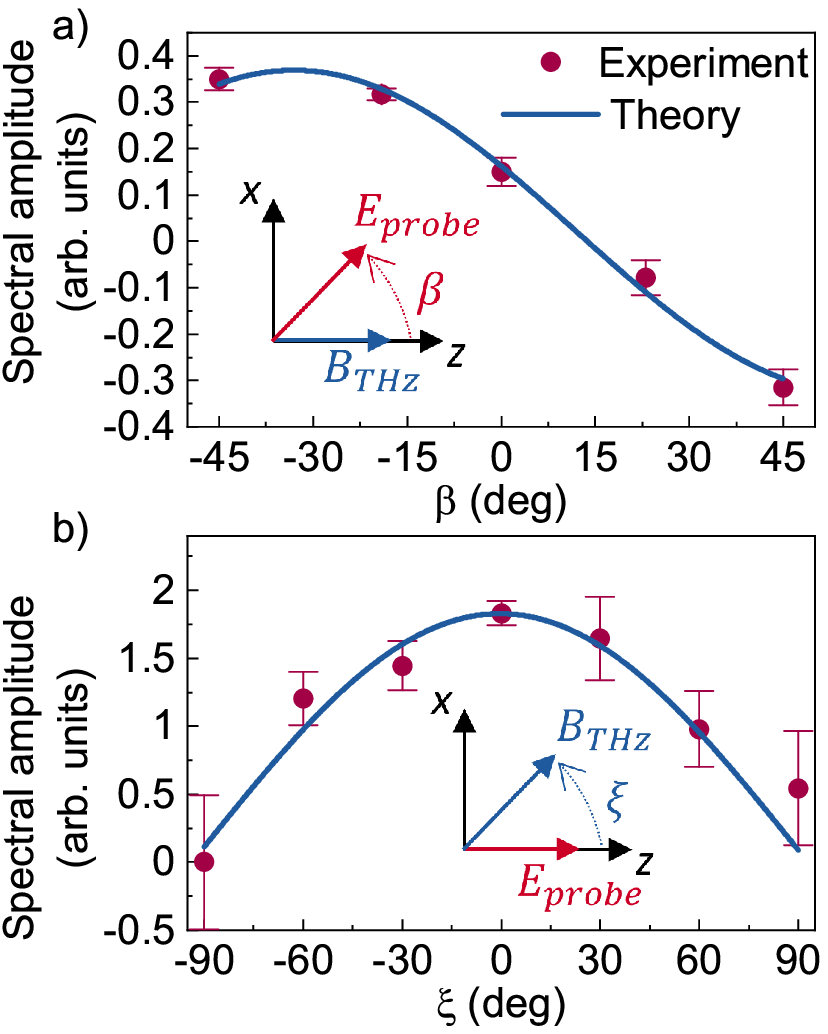}
\caption{\label{fig:polarization dependencies} The Fourier spectral amplitude of the q-AFM mode as a function of the polarization direction of the probe pulse $\beta$ for $\xi=0^\circ$ (panel a) and of the pump pulse $\xi$ for $\beta=0^\circ$ (panel b). The measurements were performed at $T=293$ K. The solid lines correspond to fits with functions given by Eqs. (\ref{eq:solution for ksi}) and (\ref{eq:A-B}), respectively.}
\end{figure}

To reveal the mechanism of excitation and detection of the mode, we systematically varied the orientation of the electric field of the THz pump $E_{THz}$ ($\xi$) and infrared probe $E_{IR}$ ($\beta$) pulses in the $x-z$ plane at $T=290$ K, i.e. in the WF phase. The observed time-traces were processed with the help of the Fourier analysis and the corresponding dependencies of the Fourier amplitudes of the q-AFM mode are shown in Fig.\ref{fig:polarization dependencies}.

The THz polarization dependence is unipolar and reaches maximum when the magnetic field of the THz pulse coincides with the direction of the external magnetic field $\mathrm{H_{ext}}$. In other words, a nonzero response in the time domain is observed when the THz magnetic field is parallel to the net magnetization in the WF phase i.e. nearly perpendicular to the slightingly canted spins.

The probe polarization dependence shows two extrema at $\beta\approx\mathrm{\pm 45^{\circ}}$ and changes the sign in the vicinity of $\beta=0^{\circ}$. Similar behaviour was observed in the polarization rotation measurements in $\mathrm{FeBO_3}$ with very similar WF spin arrangement for near-infrared light. Such a dependence evidences that the polarization ellipticity detected in the measurements originates from the magneto-optical Cotton-Mouton effect\cite{Mashkovich2019}. 

However, if the detection mechanism in the WF phase seems to be clear, much lower signal-to-noise ratio prevents us from experimental verification of the origin of the q-AFM mode below the Morin point. This is why we propose an extensive theoretical analysis of the problem.

\section{Theoretical analysis and discussion}
\subsection{Detection mechanism} 
Light matter interaction in the simplest case of electric-dipole approximation is defined by the dielectric permittivity tensor $\epsilon_{ik}$. Neglecting dissipations in the crystal it can be shown that the tensor is Hermitian ($\epsilon_{ji}=\epsilon^*_{ij}$)\cite{Landau1984}. Expansion of the tensor in a power series of $\mathbf{l}$ up to the second order terms and the subsequent diagonalization of the tensor show that light in such a crystal will have two mutually orthogonal modes with refractive indices $n_{1,2}$. In particular, for the geometry when $\mathbf{H_{ext}}\neq0$ and $\mathbf{k}\parallel c$-axis we obtain \cite{Akhmadullin2002}:
\begin{align}
    n^2_{1,2}=\frac{1}{2}\left[\left(\Omega_{xx}+\Omega_{yy}\right)\pm\sqrt{\left(\Omega_{xx}-\Omega_{yy}\right)^2+4\Omega^2_{xy}} \right],
\end{align}
\noindent where ${\Omega_{xx}=\epsilon_{xx}-\epsilon^2_{xz}/\epsilon_{zz}}$, ${\Omega_{yy}=\epsilon_{yy}-\epsilon^2_{yz}/\epsilon_{zz}}$ and ${\Omega_{xy}=\epsilon_{xy}-\epsilon_{xz}\epsilon^*_{yz}/\epsilon_{zz}}$. Note that in this section, for convenience, we use Cartesian coordinates with the $z$ axis as the optical axis ([001] crystallographic axis), the $x$ and $y$ axes lie in the sample plane and $x$ is parallel to the direction of the external magnetic field.

In the expansion of the dielectric tensor only symmetric part $\epsilon_{ij}^{(s)}=\epsilon_{ji}^{(s)}$ depends on the quadratic terms with respect to $\mathbf{l}$. This dependence eventually results in the magnetic linear birefringence and magnetic linear dichroism, which are also often called the magneto-optical Cotton-Mouton effect \cite{Ferre1984}.

In order to understand how the polarization of the IR pulse changes upon propagation, we use formalism of the Jones matrices \cite{Azzam1977}. 
First, the electrical vector of the optical pulse is rotated from the laboratory coordinate system to the coordinate system formed by the crystallographic axes. Subsequently, the static phase retardation $\Gamma$ in the sample is applied and, finally, the electric vector is returned back to the laboratory coordinate system.

By applying similar procedure we take into account the optical elements in our detection, consisting of quarter-wave plate, Wolaston prism, and balanced detector, it can be shown that the signal intensity from the differential channel of balance detector can be written in the following form
\begin{equation}
    I_{D1-D2}={E^2_{0}}\sin\left(\Gamma\right)\sin\left(2\beta\right).
    \label{Eq:differential signal from the diodes!}
\end{equation}
\noindent where $E_{0}$ is the electric field of the incident IR pulse. 

If the THz pulse is applied, the phase retardation $\Gamma$ can be represented as a sum of the time-independent and the THz-induced contributions $\Gamma=\Gamma_{0}+\delta\Gamma\left(t\right)$. Taking into account that the THz-induced signal is relatively small compared to the static one $\Gamma_0$, we obtain
\begin{equation}
\label{eq:A-B}
I_{D1-D2}\approx{E^2_{0}}\sin\left(2\beta\right)\cos\left(\Gamma_{0}\right)\delta\Gamma\left(t\right).
\end{equation}

Note, the Eq. $\ref{eq:A-B}$ gives good results for the case when the vector $\mathbf{l}$ is in the plane (WF, $T>T_M$). 
Below $T_M$, the Neel vector can be detected either due to oscillations which result in projection of $\mathbf{l}$ onto the $x-y$ plane or due to a tilt of the sample so that the probe beam propagates at a nonzero angle with respect to the $z$-axis.



\subsection{Ground state}
An equilibrium orientation of the magnetization is determined by the minimum of free energy of the crystal. The expansion of the thermodynamic potential, containing the invariants respect to the $\overline{3}m(D_{3d})$ point group of $\mathrm{\alpha-Fe_2O_3}$  in terms of normalized antiferromagnetic $\mathbf{l}$ and ferromagnetic $\mathbf{m}$ vectors, takes the following form \cite{Dzyaloshinsky1958,Turov1965,Strugatsky2015}:
\begin{align}
W=\frac{1}{2}\mathfrak{J} \mathbf{m}^2+\frac{1}{2}b_1\left(\mathbf{l}\cdot\mathbf{n}\right)^2&+\frac{1}{4}b_2\left(\mathbf{l}\cdot\mathbf{n}\right)^4\nonumber\\&+\frac{1}{2}D\left(\mathbf{m}\cdot\mathbf{l}\right)^2-2M_0\mathbf{m}\cdot\mathbf{H}_t,
    \label{eq:invariant in the general form}
\end{align}

Here $\mathfrak{J}=4H_{ex1}M_0$ and $D=4H_{ex2}M_0$ are the effective values describing the isotropic exchange interaction, $b_1=2H_{a1} M_0$ and $b_2=-2H_{a2} M_0$ are uniaxial out-of-plane anisotropy constants of the second and the fourth order, respectivetly. $\mathbf{n}$ is the direction of the easy axis. The last term ${\mathbf{H}_t=\mathbf{H}_0+\mathbf{H}_{THz}+\mathbf{d}\times\mathbf{l}}$ comprises the interactions with the static external magnetic field $\mathbf{H_0}$, the magnetic field of the THz pulse $\mathbf{H}_{THz}$ and the Dzyaloshinskii interaction, where the Dzyaloshinskii vector $\mathbf{d}$ is oriented along [001] crystallographic axis.

Looking for a minimum of the thermodynamic potential given by Eq.\ref{eq:invariant in the general form} and neglecting the longitudinal magnetic susceptibility $\chi_{||}$ of the antiferromagnet with respect to the transverse one $\chi_\perp=4M^2_0/\mathfrak{J}$ ($D\mathbf{l}^2\gg\mathfrak{J}$), one can represent $\mathbf{m}$ as a function of $\mathbf{l}$:
\begin{equation}
    \mathbf{m}=\frac{\chi_\perp}{2M_0}\left(\mathbf{H}_t-\mathbf{l}\left(\mathbf{H}_{t}\cdot\mathbf{l}\right)\right).
    \label{eq:m as a function of l}
\end{equation}
\noindent where we assumed that $|M_1|=|M_2|=M_0$ which is equivalent to the fulfillment of the conditions $\mathbf{m}^2+\mathbf{l}^2=1$ and $\mathbf{ml}=0$ \cite{Turov1965}.

Substituting (\ref{eq:m as a function of l}) in (\ref{eq:invariant in the general form}) gives the following form:
\begin{align}
    W=\frac{1}{2} b_1\left(\mathbf{l},\mathbf{n}\right)^2+\frac{1}{4}b_2\left(\mathbf{l},\mathbf{n}\right)^4-\frac{\chi_\perp}{2}\left(\mathbf{H}^2_t-\left(\mathbf{H}_t\cdot \mathbf{l}\right)^2\right).
    \label{eq:min W}
\end{align}

The value of the potential minimum depends on the mutual orientation of the vectors $\mathbf{l}$ and $\mathbf{n}$. In order to define the magnetic ground state of the crystal, it is convenient to introduce the q-AFM-vector in spherical coordinates: ${\mathbf{l}=\left(\sin{\theta}\cos{\varphi},\sin{\theta}\sin{\varphi},\cos{\theta}\right)}$. The direction of the $y$-axis coincides with the optical axis ([001] crystallographic axis), while $x$ and $z$ axes line in the sample plane and $z$ is along the two-fold symmetry axis (Fig. \ref{fig:geometry}b). 
In this coordinate system, the total energy (\ref{eq:min W}) is given by the expression:
\begin{align}
    W=\frac{1}{2} \Tilde{b}_1\sin^2{\theta}\sin^2{\varphi}+\frac{1}{4}b_2\sin^4{\theta}\sin^4{\varphi}-\frac{\chi_\perp}{2} H^2_0 \sin^2{\theta},
    \label{eq:potential_norm_form}
\end{align}
\noindent where the Dzyaloshinskii field is included in the redefined single-ion anisotropy constant ${\Tilde{b}_1=b_1-\chi_\perp \mathbf{d}^2}$. 

There are two stable states for $\theta$ and $\varphi$ that satisfy the global minimum of free energy: an easy plane (EP) and an easy axis. The EP phase corresponds to the case when $\mathbf{l}\perp c$ while for the EA phase  $\mathbf{l}|| c$

Finding the minimum value of the potential (\ref{eq:potential_norm_form}), it is important to note  that if $\mathbf{H}^2_0\ll \mathbf{d}^2$ and $b_2<0$ the stability regions for $\mathbf{l}||c$ and $\mathbf{l}\perp c$ overlap. Indeed, since $\tilde b=\tilde b\left(T\right)$, the expansion of free energy in Taylor series with powers of ${T-T_1}$ , where $T_1$ is the temperature at which the high-temperature phase loses stability, shows coexistence of two different phases:
\begin{align}
    &\mathbf{l}\perp c:\: \theta_0=\frac{\pi}{2}\quad \varphi=\:0,\:\pi \quad at \quad T>T_1,\\
    &\mathbf{l}\parallel c:\: \theta_0=\frac{\pi}{2} \quad \varphi=\frac{\pi}{2},\:\frac{3\pi}{2} \quad at \quad T<T_2=T_1\left(1+\frac{|b_2|}{b}\right).
\end{align}

Estimates of the corresponding temperatures will be given in the section below.

\subsection{Excitation mechanism} 
The oscillations dynamics and the temperature dependencies can be understood by starting out with the expressions for the effective Lagrangian $\mathcal{L}$ and the dissipative Rayleigh function $\mathcal{R}$ \cite{Andreev1980,Zvezdin1981}. In the selected spherical coordinates they take the following form:
\begin{widetext}
\begin{align}
    \mathcal{L}&=\frac{\chi_\perp}{2}\left(\left(\frac{\dot \varphi}{\gamma}-H_z\right)\sin{\theta}+\left(H_x\cos{\varphi}+H_y\sin{\varphi}\right)\cos{\theta}\right)^2+\frac{\chi_\perp}{2}\left(\frac{\dot \theta}{\gamma}-H_x\sin{\varphi}+H_y\cos{\varphi}\right)^2-\nonumber\\&\phantom{---------------------------}-\frac{1}{2} b_1\sin^2{\theta}\sin^2{\varphi}-\frac{1}{4}b_2\sin^4{\theta}\sin^4{\varphi},\\
    \mathcal{R}&=\frac{\alpha M_0}{\gamma}\left(\dot\theta^2+\sin^2{\theta}\dot\varphi^2\right)
\end{align}
\end{widetext}
\noindent where $\gamma$ is the gyromagnetic ratio, $\alpha$ is the Gilbert dumping constant. The $H_x$, $H_y$, $H_z$ are the components of ${\mathbf{H}_t=(d\cos\theta, 0, H_0 - d\sin\theta\cos\varphi)}$.

Above the Morin point, i.e. in the easy plane phase ($T>T_M$), the linearized Lagrange-Euler equations ${\frac{d}{dt}\left( \frac{\partial \mathcal{\mathcal{L}}}{\partial \dot{\theta}}\right) - \frac{\partial \mathcal{L}}{\partial \theta}+\frac{\partial \mathcal{R}}{\partial \theta}=0}$, etc. with the angles ${\theta=\theta_1+\pi/2}$ and ${\varphi=\varphi_1+0}$ take the following form:

\begin{align}
\ddot\theta_1 +  \frac{2}{\tau} \dot\theta_1 + \gamma^2(H_0^2 - H_0d)\theta_1 &= 0,  \label{eq:EP_motion_equations1}\\ 
\ddot\varphi_1 +  \frac{2}{\tau} \dot\varphi_1 + \gamma^2\left(d(d-H_0) +\frac{b_1}{\chi_\perp} \right) \varphi_1 &= \gamma \dot H_z
\label{eq:EP_motion_equations2}
\end{align}
\noindent where $\tau$ is a damping constant \cite{Zvezdin1979,Zvezdin1981}.

A characteristic property of dynamic equations \ref{eq:EP_motion_equations1} and \ref{eq:EP_motion_equations2} is that the driving force is proportional to $\mathbf{\dot H}$. This effect was first predicted theoretically in Refs. 
\cite{Zvezdin1979, Zvezdin1981}, and then later discovered for IR light in NiO \cite{Satoh2010}.

If we identify $H_D \equiv -d$, $H^2_c=2H_EH_{a1}$ and take into account that $d(d-H_0) \approx d^2$, we recover the traditional notations for q-AFM and q-FM frequencies:
\begin{eqnarray}
\omega^2_{AFM} &\approx& \gamma^2 \left(H^2_c+H_D\left(H_0+H_D\right)\right), \label{eq:AFM_frequency}\\
\omega^2_{FM}  &=&  \gamma^2\left(H_0^2 + H_0H_D\right).
\end{eqnarray}

It is seen that the equations of motion (\ref{eq:EP_motion_equations1}) and (\ref{eq:EP_motion_equations2}) have a form typical for a damped harmonic oscillator where the magneto-dipole (Zeeman) interaction between the spins and the magnetic field of the THz pulse acts as a toque for the q-AFM mode, while the torque for the q-FM mode is equal to zero in the selected experimental geometry. Considering only the first term in the Fourier expansion of the magnetic field of the THz pulse ${\mathbf{H}_{THz}=H_\perp\left(t\right)\left(-\sin{\xi}, 0, \cos{\xi}\right)}$, which is responsible for resonant excitation,the solution takes the form:
\begin{align}
\begin{matrix}
    \varphi_1\left(t\right) =-\gamma H^{\omega_{AFM}}_\perp\cos{\omega_{AFM} t}e^{-t/\tau}
    \cos{\xi}
\end{matrix}
\label{eq:solution for ksi}
\end{align}

The approximation of the THz polarization dependence with the Eq.\ref{eq:solution for ksi} shows good agreement with the experimental data in the high temperature ($T>T_M$) phase  (Fig.\ref{fig:polarization dependencies}b)).

The dependence of $H^2_{c}$ is linear with the temperature ${H^2_{c}=aT+p}$. Then, substituting the values for $a=67\cdot10^3$ $\mathrm{kOe^2\mathtt{/}deg}$, $p=-16.9\cdot 10^3$ $\mathrm{kOe^2}$ and $H_D=22$ kOe from {Ref. [\onlinecite{Velikov1969}]}, the fit with Eq.\ref{eq:AFM_frequency} shows very good agreement with the experimental data (Fig. \ref{fig:temp_dep}c, blue solid line).

Repeating the linearization procedure for the easy axis phase ($T<T_M$, $\theta =\theta_1+\pi/2$ and $\varphi = \varphi_1+\pi/2$), we obtain:
\begin{align}
    \ddot\theta_1 + \frac{2}{\tau}\dot\theta_1-\frac{\gamma^2}{\chi_\perp}\left(b_1+b_2- 2\chi_\perp H^2_0\right)\theta_1=\gamma\dot H_x,
    \label{eq:EA_motion_equations1}
    \\
    \ddot\varphi_1+ \frac{2}{\tau}\dot\varphi_1-\frac{\gamma^2}{\chi_\perp}\left(b_1+b_2\right)\varphi_1=\gamma\dot H_z,
    \label{eq:EA_motion_equations2}
\end{align}

Returning to the defenitions of the exchange and the anisotropy fields and neglecting the external magnetic field with respect to the anisotropy field, we observe the case of two degenerate modes of antiferromagnetic resonance with the frequency:
\begin{align}
    \omega_{AFM}^2=\gamma^2H_E\left(H_{a2}-H_{a1}\right).
    \label{eq:AFM_frequency_LowT}
\end{align}

As in the previous case, the equations of motion are equations for harmonically damped oscillators and the q-AFM mode can be triggered with a Zeeman torque. Thereby, the solution is as follows:
\begin{align}
\begin{matrix}
    \begin{pmatrix}
    \theta_1\left(t\right)\\
    \varphi_1\left(t\right)
    \end{pmatrix} &
    =-\gamma H^{\omega_{AFM}}_\perp\cos{\omega_i t}e^{-t/\tau}
    \begin{pmatrix}
    -\sin{\xi}\\
    \cos{\xi}
    \end{pmatrix}
\end{matrix}
\label{eq:solution lowT}
\end{align}
\noindent where $H^{\omega_{AFM}}_\perp$ is a spectral component of the magnetic field of the THz pulse and $\omega$ is the frequency of the AFM mode in the easy axis phase.

The fit of the temperature dependence below the Morin temperature ($T<T_M$) using the Eq. \ref{eq:AFM_frequency_LowT} shows a good agreement with the experimental data in the vicinity of $T_M$ (Fig. \ref{fig:temp_dep}, blue solid line).

An extension of the theory to the low temperature range can be obtained with the help of a spin-wave model\cite{Nagai1975,Chou2012}. The fit with this model is shown as a green dash dot line in Fig. \ref{fig:temp_dep}c). However, despite a good match one should remember that the spin-wave model simply has more fit parameters.

Finally, we can estimate the values for the temperatures $T_1$ and $T_2$ at which the WF and the AF phases lose stability. The temperatures $T_1$ and $T_2$ can be obtained by extrapolation of the temperature dependence of the frequency of the q-AFM mode given by Eqs.\ref{eq:AFM_frequency} and \ref{eq:AFM_frequency_LowT} to the intersections with the horizontal axis (see Fig.\ref{fig:temp_dep}c).
The intersection of the two temperature dependencies gives the value of the Morin temperature $T_M$. At this point the WF and AFM phases have equal free energies ${W_{WF}=W_{AFM}}$. Thus, substituting the values $b_1=200$ Oe and ${b_2/b_1=-0.23}$ \cite{Levitin1968}, from Eqs.\ref{eq:AFM_frequency} and \ref{eq:AFM_frequency_LowT} we obtain the following estimates:  $T_1\approx235$ K, $T_2\approx282$, $T_M\approx258$ K.
The obtained Morin transition temperature is close to the values reported earlier \cite{Morin1950,Dzyaloshinsky1958,Velikov1969}.

\section{Conslusions}
We experimentally demonstrated that ultrafast dynamics triggered in hematite $\mathrm{\alpha-Fe_{2}O_{3}}$ by intense nearly single-cycle terahertz pulses. We show that the dynamics corresponds to the quasi-antiferromagnetic resonant mode in the compound. The mode is detected due to the magneto-optical Cotton-Mouton effect and excited employing mechanism of magnetic dipole interaction of THz magnetic field with spins. The coupling efficiency is proportional to the time derivative of the magnetic field of the THz pulse and reaches the maximum when the orientation of the THz magnetic field is perpendicular to the antiferromagnetically coupled spins. Using theoretically derived expressions we fitted the experimentally observed temperature dependencies of the mode near the Morin point and estimated the temperatures at which the high and low temperature phases lose stability.

\begin{acknowledgments}
The authors thank S. Semin and C. Berkhout for technical support. This work was supported by de Nederlandse organisatie voor wetenschappelijk onderzoek (NWO) and the Russian Foundation for Basic Research (grant 18-02-40027). A.K.Z. acknowledges the financial support from the Russian Science Foundation (grant 17-12-01333). 
\end{acknowledgments}
\bibliography{bibliography}
\end{document}